\documentclass{osa-article}
\journal{oe}

\usepackage{amsmath}
\usepackage{bm}
\usepackage{physics}

\newcommand{\mup}{\ensuremath{\mu_\text{p}}}
\newcommand{\sigalpha}{\ensuremath{\sigma_\alpha}}
\newcommand{\sigbeta}{\ensuremath{\sigma_\beta}}
\newcommand{\sigtau}{\ensuremath{\sigma_\tau}}
\newcommand{\sighatalpha}{\ensuremath{\hat{\sigma}_\alpha}}
\newcommand{\sighatbeta}{\ensuremath{\hat{\sigma}_\beta}}
\newcommand{\sighattau}{\ensuremath{\hat{\sigma}_\tau}}

\newcommand{\transpose}{\ensuremath{\mathsf{T}}}

\newcommand{\conv}{\ensuremath{*}}
\newcommand{\circconv}{\ensuremath{\circledast}}
\newcommand{\psd}{\ensuremath{S_{xx}}}

\newcommand{\rvec}{\ensuremath{\bm{r}}}
\newcommand{\xvec}{\ensuremath{\bm{x}}}
\newcommand{\yvec}{\ensuremath{\bm{y}}}
\newcommand{\Avec}{\ensuremath{\bm{A}}}
\newcommand{\muvec}{\ensuremath{\bm{\mu}}}
\newcommand{\psivec}{\ensuremath{\bm{\psi}}}
\newcommand{\lambdavec}{\ensuremath{\bm{\lambda}}}
\newcommand{\thetavec}{\ensuremath{\bm{\theta}}}
\newcommand{\etavec}{\ensuremath{\bm{\eta}}}
\newcommand{\zetavec}{\ensuremath{\bm{\zeta}}}

\newcommand{\Vmat}{\ensuremath{\bm{V}}}
\newcommand{\Umat}{\ensuremath{\bm{U}}}
\newcommand{\hmat}{\ensuremath{\bm{h}}}
\newcommand{\Dmat}{\ensuremath{\bm{D}}}
\newcommand{\Imat}{\ensuremath{\bm{I}}}
\newcommand{\Sigmax}{\ensuremath{\bm{\Sigma_x}}}
\newcommand{\Sigmay}{\ensuremath{\bm{\Sigma_y}}}

\newcommand{\Xmat}{\ensuremath{\bm{X}}}
\newcommand{\Zmat}{\ensuremath{\bm{Z}}}
\newcommand{\Smat}{\ensuremath{\bm{S}}}

\newcommand{\muhat}{\ensuremath{\hat{\mu}}}
\newcommand{\psihat}{\ensuremath{\hat{\psi}}}
\newcommand{\lambdahat}{\ensuremath{\hat{\lambda}}}
\newcommand{\thetahat}{\ensuremath{\hat{\theta}}}

\newcommand{\muhatvec}{\ensuremath{\bm{\hat{\mu}}}}
\newcommand{\psihatvec}{\ensuremath{\bm{\hat{\psi}}}}
\newcommand{\lambdahatvec}{\ensuremath{\bm{\hat{\lambda}}}}
\newcommand{\thetahatvec}{\ensuremath{\bm{\hat{\theta}}}}

\DeclareMathOperator{\Cov}{Cov}

\DeclareMathOperator{\AIC}{AIC}

\begin{document}
\title{Maximum-likelihood parameter estimation in terahertz time-domain spectroscopy}

\author{Laleh Mohtashemi, Paul Westlund, Derek G.~Sahota, Graham B.~Lea, Ian Bushfield, Payam Mousavi, and J.~Steven Dodge\authormark{*}}

\address{Department of Physics, Simon Fraser University, V5A~1S6, Canada}

\email{\authormark{*}jsdodge@sfu.ca}

\begin{abstract}
We present a maximum-likelihood method for parameter estimation in terahertz time-domain spectroscopy. We derive the likelihood function for a parameterized frequency response function, given a pair of time-domain waveforms with known time-dependent noise amplitudes. The method provides parameter estimates that are superior to other commonly-used methods, and provides a reliable measure of the goodness of fit. We also develop a simple noise model that is parameterized by three dominant sources, and derive the likelihood function for their amplitudes in terms of a set of repeated waveform measurements. We demonstrate the method with applications to material characterization.
\end{abstract}

\section{Introduction}
\label{sec:Introduction}

At the heart of most applications of terahertz time-domain spectroscopy (THz-TDS) is a mathematical procedure that relates raw THz-TDS waveform measurements to parameters of scientific and technological interest~\cite{duvillaret1999, dorney2001, naftaly2015}. Typically this analysis is formulated in the frequency domain, since it provides the most natural description of any linear, time-invariant system of interest. But since a THz-TDS waveform describes the electric field as a function of \emph{time}, it must be Fourier-transformed for use in any frequency-domain analysis. The Fourier transform scrambles the THz-TDS measurement noise, which is more naturally represented in the time domain, and produces artifacts that can degrade the overall measurement quality and yield misleading results\cite{withayachumnankul2008a, naftaly2009, naftaly2013, naftaly2017, skorobogatiy2018, hubers2011, kruger2011, jahn2016, rehn2017, withayachumnankul2014}. Furthermore, the standard approaches to parameter estimation in THz-TDS involve the ratio of two noisy waveforms, which further distorts the noise spectrum and can yield noise distributions that are far from Gaussian. Other approaches have emerged that improve on the standard procedures~\cite{vanmechelen2014, krimi2016, peretti2019}, but so far all of the existing approaches to THz-TDS analysis lack clear, model-independent statistical criteria for establishing parameter confidence intervals or for deciding whether a given physical model describes the data well or not. Here, we introduce a general framework to remedy this~\cite{mohtashemi2020}. We describe a maximum-likelihood approach to THz-TDS analysis in which both the signal and the noise are treated explicitly in the time domain, together with a frequency-domain constraint between the input and the output signal. We show that this approach produces superior results to existing analysis methods.

\section{Signal and noise in THz-TDS}
\label{sec:signal-noise}
\begin{figure}[tbp]
\begin{center}
\includegraphics[width=0.9\columnwidth]{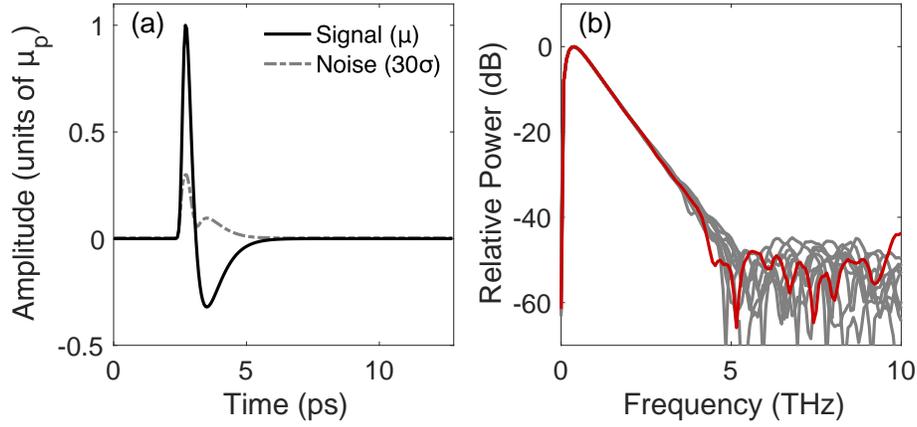}
\caption{(a) Simulated time-domain signal (black solid line) and noise amplitude (gray dashed line). (b) Power spectral density (relative to peak) obtained from ten simulated time-domain measurements, with one shown in red and the rest shown in gray.}
\label{fig:signal-noise}
\end{center}
\end{figure}

A Monte Carlo simulation of an elementary THz-TDS analysis illustrates the basic problems that our method solves; it also allows us to develop notational conventions that we will use to describe our main results. Figure~\ref{fig:signal-noise}(a) shows an ideal, noiseless THz-TDS waveform $\mu(t)$ normalized to its peak value $\mup$~\cite{grischkowsky1991}, together with a time-domain Gaussian noise amplitude $\sigma_\mu(t)$ given by
\begin{equation}
\sigma_\mu^2(t) = \sigalpha^2 + [\sigbeta\mu(t)]^2 + \left[\sigtau\dot{\mu}(t)\right]^2
\label{eq:noise-amplitude}
\end{equation}
with amplitudes $\sigalpha = 10^{-4}\mup$, $\sigbeta = 10^{-2}$, and $\sigtau = 1~\text{fs}$. Physically, the additive noise term $\sigalpha$ is produced by the detection electronics (in units of $\mup$); the multiplicative noise term $\sigbeta|\mu(t)|$ is produced by laser power fluctuations, which modulate the signal strength; and the jitter term $\sigtau|\dot{\mu}(t)| \approx |\mu(t + \sigtau)-\mu(t)|$ is produced by fluctuations in the optical path lengths that control the delay between the terahertz pulse arrival time and the receiver sampling time. In this example the overall noise is dominated by $\sigbeta|\mu(t)|$, which produces the two peaks near $t=2.75~\text{ps}$ and $t=3.50~\text{ps}$. The contribution from $\sigtau|\dot{\mu}(t)|$ is less evident, except at $t=3.10~\text{ps}$, where $\mu(t)$ crosses zero with a nonzero slope. The contribution from $\sigalpha$ dominates at times when both $\mu(t)$ and $\dot{\mu}(t)$ are small, although in the figure it is barely discernible from zero. Such structured, signal-dependent, time-domain noise is common in THz-TDS waveform measurements, and leads to well-known ambiguities in defining the signal-to-noise ratio and dynamic range for them~\cite{naftaly2009, naftaly2013, skorobogatiy2018}.

We simulate a single THz-TDS waveform measurement by computing 
\begin{equation}
x(t_k) = \mu(t_k) + \sigma_\mu(t_k)\epsilon(t_{k})
\label{eq:measured-waveform}
\end{equation}
at equally spaced times $t_k = kT, k \in \{0, 1,\ldots, N-1\}$, with $N = 256$ and $T = 50~\text{fs}$, where each $\epsilon(t_k)$ is an independent random variable with a standard normal distribution. This sequence has the discrete Fourier transform (DFT)
\begin{equation}
X(\omega_l) = \tilde{x}(\omega_l) = \tilde{\mu}(\omega_l) + [\tilde{\sigma}_\mu\circconv\tilde{\epsilon}](\omega_l)
\label{eq:measured-spectrum}
\end{equation}
at the discrete frequencies $\omega_l = 2\pi l/(NT), l \in \{0, 1, \ldots, N-1\}$, where $\tilde{x}(\omega_l)$, $\tilde{\sigma}_\mu(\omega_l)$, and $\tilde{\epsilon}(\omega_l)$ denote the DFTs of $x(t_k)$, $\sigma_\mu(t_k)$, and  $\epsilon(t_k)$, respectively, and the $\circconv$ symbol denotes circular discrete convolution. Figure~\ref{fig:signal-noise}(b) shows the power spectral density $\psd(\omega_l) = (T/N)|X(\omega_l)|^2$ at the unique frequencies given by $l \leq \operatorname{floor}(N/2)=\left\lfloor N/2\right\rfloor$ for ten such simulations, where each spectrum is normalized to its peak value. The signal decreases exponentially with frequency, falling by a bit more than 40~dB from its peak power before reaching the noise floor near 5~THz. The red-highlighted spectrum in Fig.~\ref{fig:signal-noise}(b) shows that while the noise is relatively flat between 5~THz and 10~THz, it exhibits oscillatory fluctuations with a period of about 0.75~THz. These also appear in all of the other spectra in Fig.~\ref{fig:signal-noise}(b), and arise because the convolution $[\tilde{\sigma}_\mu\circconv\tilde{\epsilon}](\omega_l)$ smooths the noise over a frequency scale given by the inverse width of $\sigma_\mu(t)$. The resulting correlations blur the distinction between the signal and the noise, and their presence implies that the true uncertainty in $X(\omega_l)$ is significantly smaller than the noise floor suggested by $\psd(\omega_l)$, which neglects the information that noise at one frequency provides about the noise at neighbouring frequencies.

\section{Transfer function estimation in THz-TDS}
\label{sec:transfer-function-estimation}
To measure the response of a system requires measurements of two THz-TDS waveforms: an input, $\mu(t)$, and an output, $\psi(t) = [h\conv\mu](t)$, where $h(t)$ is the impulse response of the system and the $\conv$ symbol denotes continuous-time convolution. Fourier transforming the ideal relationship $\psi(t) = [h\conv\mu](t)$ converts the time-domain convolution to a frequency-domain multiplication, $\tilde{\psi}_\text{c}(\omega) = H(\omega)\tilde{\mu}_\text{c}(\omega)$, where $\tilde{\mu}_\text{c}(\omega)$ and $\tilde{\psi}_\text{c}(\omega)$ denote the continuous-time Fourier transforms of $\mu(t)$ and $\psi(t)$, respectively, and $H(\omega) = \tilde{h}_\text{c}(\omega)$ is the continuous-time transfer function of the system. Proceeding as we did for a single waveform, we simulate input and output waveform measurements $x(t_k)$ and $y(t_k)$, respectively, by computing $\mu(t)$ and $\psi(t)$ at the discrete times $t_k$ and adding time-domain noise $\sigma_\mu(t_k)\epsilon(t_{k})$ and $\sigma_\psi(t_k)\delta(t_k)$, where each $\epsilon(t_k)$ and $\delta(t_k)$ is an independent random variable with a standard normal distribution. After applying the discrete Fourier transform to obtain $X(\omega_l) = \tilde{x}(\omega_l)$ and $Y(\omega_l) = \tilde{y}(\omega_l)$, we can determine the \emph{empirical transfer function estimate} (ETFE)~\cite{ljung1999},
\begin{equation}
\hat{H}_\text{ETFE}(\omega_l) = \frac{Y(\omega_l)}{X(\omega_l)} = \frac{\tilde{\psi}(\omega_l) + [\tilde{\sigma}_\psi\circconv\tilde{\epsilon}](\omega_l)}{\tilde{\mu}(\omega_l) + [\tilde{\sigma}_\mu\circconv\tilde{\delta}](\omega_l)},
\label{eq:ETFE}
\end{equation}
where $\tilde{\mu}(\omega_l)$ and $\tilde{\psi}(\omega_l)$ are the DFTs of the ideal input and output signals, respectively, that we would obtain in  the absence of noise, and  $[\tilde{\sigma}_\mu\circconv\tilde{\epsilon}](\omega_l)$ and $[\tilde{\sigma}_\psi\circconv\tilde{\delta}](\omega_l)$ are the corresponding frequency-domain noise terms. The ETFE is a common starting point for THz-TDS analysis---it is easy to compute from the raw data, and it provides an estimate of $H(\omega)$ for any linear, time-invariant system. Frequently, though, the focus of interest is not $H(\omega)$ itself but a relatively small number of parameters that specify $H(\omega)$ through a physical model, such as the thickness and refractive index of a material. In this case, fitting the model directly to $\hat{H}_\text{ETFE}(\omega_l)$ can yield ambiguous results, because $\hat{H}_\text{ETFE}(\omega_l)$ is biased and has noise that is both correlated and non-Gaussian.

\begin{figure}[tbp]
\begin{center}
\includegraphics[width=0.9\columnwidth]{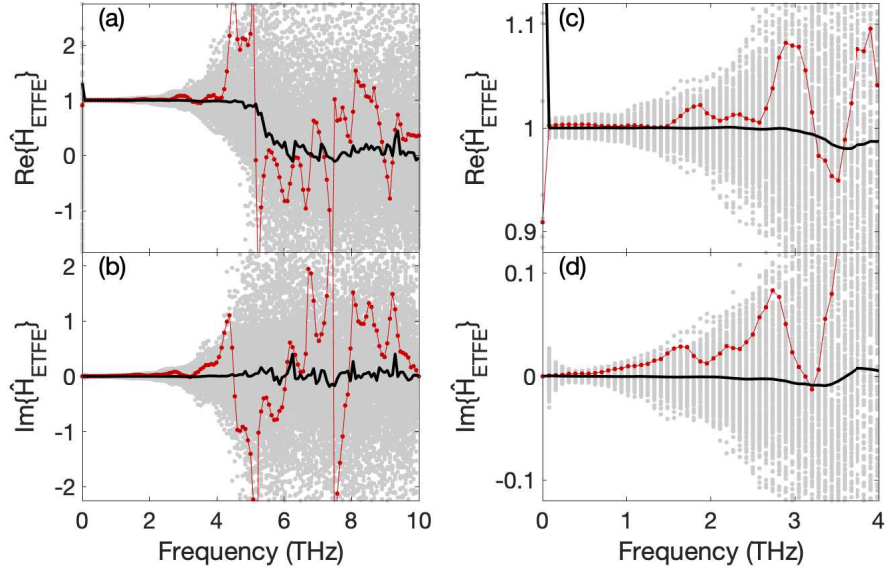}
\caption{Gray dots show the real (a,c) and imaginary parts (b,d) of the empirical transfer function estimate $\hat{H}_\text{ETFE}$, Eq.~(\ref{eq:ETFE}), for 250 pairs of simulated time-domain measurements of the waveform shown in Fig.~\ref{fig:signal-noise}(a). One estimate is highlighted in red, with the dots connected by a thin red line. The solid black line shows the average over all simulations. Panels (a,b) show the full bandwidth and (c,d) show the same data over the primary signal bandwidth.}
\label{fig:ETFE-simulation}
\end{center}
\end{figure}

To illustrate these deficiencies, Fig.~\ref{fig:ETFE-simulation} shows 250 simulations of $\hat{H}_\text{ETFE}(\omega_l)$ with nominally identical input and output pulses, corresponding to $H(\omega) = 1$. The red-highlighted curves show correlations similar to those shown in Fig.~\ref{fig:signal-noise}, but that extend to frequencies where the signal is much stronger. The average over all simulations reveals the bias in $\hat{H}_\text{ETFE}(\omega_l)$: in Fig.~\ref{fig:ETFE-simulation}(a), $\Re\{\hat{H}_\text{ETFE}\}$ falls from the correct value of 1 to the biased value of 0 as the frequency increases above 5~THz, where the signal reaches the noise floor in Fig.~\ref{fig:signal-noise}(b). Also, the noise distribution for $\hat{H}_\text{ETFE}(\omega_l)$ develops wide, non-Gaussian tails at frequencies where the signal is small, because noise fluctuation that cause $|X(\omega_l)|\rightarrow 0$ become more likely, and these in turn cause $\hat{H}_\text{ETFE}(\omega_l)$ to diverge~\cite{guillaume1996}. If the noise is uncorrelated, such large fluctuations are only expected when the signal is small. But in the more typical case that the noise is correlated, they can influence frequencies well within the primary signal bandwidth, as indicated by the red-highlighted curves in Figs.~\ref{fig:ETFE-simulation}(c) and \ref{fig:ETFE-simulation}(d).

\begin{figure}[tbp]
\begin{center}
\includegraphics[width=0.9\columnwidth]{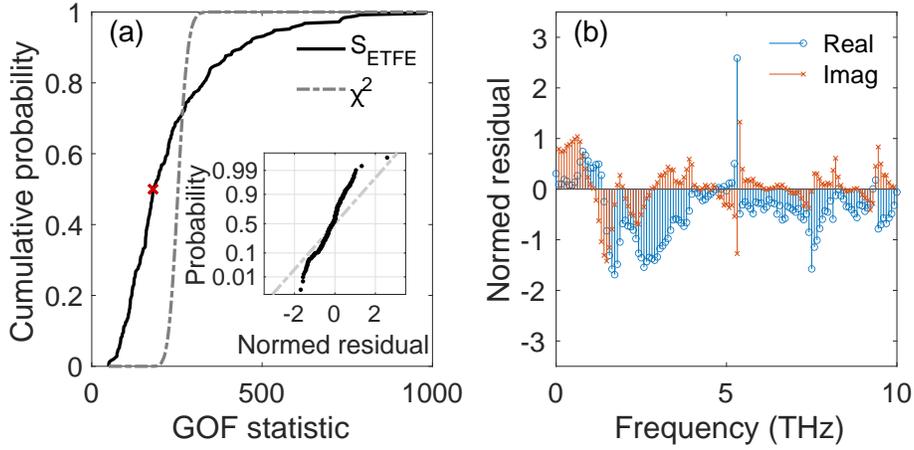}
\caption{Measures of fit quality for ETFE fits with Eq.~(\ref{eq:model1}), obtained by minimizing Eq.~(\ref{eq:ETFE-cost}). (a)~Cumulative distribution of the GOF statistic, $S_\text{ETFE}$, for the Monte Carlo simulations shown in Fig.~\ref{fig:ETFE-simulation}. The $\chi^2$ distribution for the same number of degrees of freedom ($\nu = 254$) is shown for comparison. The red $\times$ marks a fit with $S_\text{ETFE} \approx 180$, which is just above the median value. The normalized residuals for this fit, $r_\text{ETFE}(\omega_l; \thetahatvec)$, are shown in (b) as a function of frequency, and in the inset to (a) with a normal probability plot (black dots, which represent both the real and the imaginary parts of $r_\text{ETFE}$). The gray dash-dotted line in the inset represents the standard normal distribution. }
\label{fig:ETFE-GOF}
\end{center}
\end{figure}

Figure~\ref{fig:ETFE-GOF} shows how these problems distort standard measures of fit quality. It displays the results of weighted least-squares fits to the ETFE simulations in Fig.~\ref{fig:ETFE-simulation} with the model 
\begin{equation}
H_1(\omega; \thetavec) = \theta_1\exp(i\omega\theta_2),
\label{eq:model1}
\end{equation}
which rescales the input by an amplitude $\theta_1$ and shifts it by a delay $\theta_2$ when we adopt the $\exp(-i\omega t)$ convention for harmonic time dependence. For each fit, the estimated parameter vector $\thetahatvec$ is chosen to minimize
\begin{equation}
Q_\text{ETFE}(\thetavec) = \sum_{l = 0}^{N-1}\frac{\left|\hat{H}_\text{ETFE}(\omega_l) - H_1(\omega_l;\thetavec)\right|^2}{\sigma_{\omega_l}^2},
\label{eq:ETFE-cost}
\end{equation}
where each $\sigma_{\omega_l}^2 = (\text{Var}\{\Re[\hat{H}_\text{ETFE}(\omega_l)]\} +\text{Var}\{\Im[\hat{H}_\text{ETFE}(\omega_l)]\})$ is determined from the Monte Carlo samples. These fits return accurate estimates for the model parameters---over all simulated samples, we obtain $\hat{\theta}_1= 1.000\pm 0.005$ and $\hat{\theta}_2 = (0.0\pm 0.8)~\text{fs}$, which are consistent with the true values, $\theta_1=1$ and $\theta_2=0$, used in the simulation. But the normalized fit residuals, given by
\begin{equation}
r_\text{ETFE}(\omega_l; \thetahatvec) =  \frac{\hat{H}_\text{ETFE}(\omega_l) - H_1(\omega_l;\thetahatvec)}{\sigma_{\omega_l}},
\label{eq:ETFE-residuals}
\end{equation}
show strong deviations from the standard normal distribution, exhibit structure that could easily be mistaken for real physical features, and yield a goodness of fit (GOF) statistic $S_\text{ETFE} = Q_\text{ETFE}(\thetahatvec)$ that looks nothing like the $\chi^2$ distribution that we would normally use to evaluate the fit quality. Also, the uncertainty estimates, $\hat{\sigma}_{\theta_1} = 0.001$ and $\hat{\sigma}_{\theta_2} = 0.2~\text{fs}$, obtained by the usual method of inverting the curvature matrix of $Q_\text{ETFE}(\thetavec)$ around the minimum of each fit, significantly underestimate the values that are actually observed over the set of Monte Carlo samples, $\sigma_{\theta_1} = 0.005$ and $\sigma_{\theta_2} = 0.8~\text{fs}$. In short, while an ETFE fit with Eq.~(\ref{eq:ETFE-cost}) may provide accurate parameter estimates when the underlying model is known in advance, it offers poor discrimination between good models and bad ones, and it yields parameter uncertainty estimates that are unrealistically precise.

An alternative approach is to use a least-squares (LS) procedure that minimizes the sum of the squared differences between the output signal and the transformed input signal~\cite{vanmechelen2014,krimi2016,peretti2019},
\begin{align}
Q_\text{LS}(\thetavec) &= \sum_{l=0}^{N-1}\left|Y(\omega_l) - H(\omega_l;\thetavec)X(\omega_l)\right|^2 = \sum_{k=0}^{N-1}\left\{y(t_k) - [h(\thetavec)\circconv x](t_k)\right\}^2.
\label{eq:LS-cost}
\end{align}
Here, $h(t_k; \thetavec)$ is the model impulse response, equal to the inverse DFT of $H(\omega_l;\thetavec)$, and $[h(\thetavec)\circconv x](t_k)$ represents the convolution of impulse response with the input vector at the time $t_k$. The equivalence between the frequency-domain sum and the time-domain sum is assured by Parseval's theorem. The LS procedure avoids the division by $X(\omega_l)$ that distorts the noise distribution of the ETFE in the small-signal limit, and the time-domain residuals $r_\text{LS}(t_k; \thetahatvec) = y(t_k) - [h(\thetavec)\circconv x](t_k)$ are a sensitive indicator of fit quality. But the uniform weighting in Eq.~(\ref{eq:LS-cost}) implicitly assumes that this noise is constant in time, which as we have noted is frequently not the case.

In principle, we should be able to account for time-dependent noise by assigning appropriate weights to the sum, but this is not as simple as it might seem. The problem is that Eq.~(\ref{eq:LS-cost}) treats the input and output noise asymmetrically, which we can see more clearly if we express it explicitly in terms of the time-domain signal and noise sequences:
\begin{equation}
Q_\text{LS}(\thetavec) = \sum_{k=0}^{N-1}\left[\psi(t_k) - [h(\thetavec)\circconv\mu](t_k) + \sigma_\psi(t_k) - [h(\thetavec)\circconv(\sigma_\mu\epsilon)](t_k)\right]^2.
\label{eq:LS-cost-noise}
\end{equation}
This asymmetry results from the implicit assumption that all noise is restricted to the output variable in an LS fit, so that the term $[h(\thetavec)\circconv(\sigma_\mu\epsilon)](t_k)$ can be ignored. When input noise is present---as it always is in THz-TDS---the optimal fit weights will depend not just on the measurement noise sequences but also on $h(t_k; \thetavec)$. Any modification of Eq.~(\ref{eq:LS-cost}) with constant weights will generically bias the parameter estimates toward values that minimize the input noise contribution, and will cause the GOF statistic, $S_\text{LS} = Q_\text{LS}(\thetahatvec)$, to have a distribution that also depends on $h(t_k; \thetavec)$. As we discuss below, these problems can all be solved with a fit procedure derived from the maximum-likelihood principle.

\section{Maximum-likelihood estimation of a parameterized transfer function model}
\label{sec:MLE-TF}

The likelihood function is equivalent to the probability density for obtaining the measured data under the assumptions of a given model, and forms the theoretical basis for the standard least-squares fit procedure. To define it in the THz-TDS context, we express the measured input and output signals in vector notation as $\xvec = [x(t_0), x(t_1), \ldots, x(t_{N-1})]^\transpose$ and $\yvec = [y(t_0), y(t_1), \ldots, y(t_{N-1})]^\transpose$, respectively, and assume that they represent noisy measurements of bandlimited but otherwise unknown ideal signal vectors $\muvec = [\mu(t_0), \mu(t_1), \ldots, \mu(t_{N-1})]^\transpose$ and $\psivec = [\psi(t_0), \psi(t_1), \ldots, \psi(t_{N-1})]^\transpose$ with noise covariance matrices $\Sigmax$ and $\Sigmay$, respectively. We further assume that $\muvec$ and $\psivec$ satisfy the linear constraint $\psivec = \hmat(\thetavec)\muvec$, where $\hmat$ is a known matrix function of an unknown $p$-dimensional parameter vector $\thetavec$. The likelihood function is then a product of two multivariate Gaussian distributions,
\begin{equation}
L(\thetavec, \muvec, \psivec; \xvec, \yvec) = \frac{\exp[-\frac{1}{2}(\xvec - \muvec)^\transpose\,\Sigmax^{-1}(\xvec-\muvec)]}{\sqrt{(2\pi)^N\det(\Sigmax)}}\frac{\exp[-\frac{1}{2}(\yvec - \psivec)^\transpose\,\Sigmay^{-1}(\yvec-\psivec)]}{\sqrt{(2\pi)^N\det(\Sigmay)}},
\label{eq:likelihood-tf}
\end{equation}
where the dependence on $\thetavec$ on the right side of the expression is implicit in the constraint between $\muvec$ and $\psivec$. The signal vectors $\muvec$ and $\psivec$ appear as arguments in the likelihood function but are otherwise unimportant---the statistical literature aptly describes these as \emph{nuisance parameters}, which we must eliminate to estimate $\thetavec$, the parameters of interest~\cite{pawitan2001}.

We can use discrete-time Fourier analysis to obtain explicit expressions for  $\hmat(\thetavec)$, $\Sigmax$, and $\Sigmay$. The time-domain transfer matrix $\hmat(\thetavec)$ is
\begin{equation}
h_{jk}(\thetavec) = \frac{1}{N}\sum_{l = -N/2 + 1}^{N/2-1} H(\omega_l; \thetavec) \exp[-2\pi i(j - k)l/N]
+ \frac{1}{N}\Re[H(\omega_{N/2};\thetavec)]\cos[\pi(j-k)],
\label{eq:transfer-matrix}
\end{equation}
where $\omega_l$ now extends to negative values of $l$, and $N$ is assumed even; for odd $N$, the sum runs from $l = -(N-1)/2$ to $(N-1)/2$ and the anomalous term at the Nyquist frequency is absent. Similarly, we can define the discrete-time matrix representation of the time-derivative operator, $\Dmat$, as
\begin{equation}
D_{jk} = \frac{1}{N}\sum_{l = -\lfloor (N-1)/2 \rfloor}^{\lfloor (N-1)/2 \rfloor} -i\omega_l \exp[-2\pi i(j - k)l/N],
\label{eq:D-matrix}
\end{equation}
by recognizing that the transfer function for the continuous time-derivative operator is $H(\omega) = -i\omega$. Note that Eq.~(\ref{eq:D-matrix}) is valid for all values of $N$ because $\Re[-i\omega_{N/2}]=0$, so the anomalous term in Eq.~(\ref{eq:transfer-matrix}) is zero. Equation~(\ref{eq:D-matrix}) allows us to express the noise variance function $\sigma_\mu^2(t)$ in Eq.~(\ref{eq:noise-amplitude}) as a matrix function in discrete time,
\begin{equation}
\left[\Vmat(\muvec)\right]_{jk} = \delta_{jk}\left[\sigalpha^2 + \sigbeta^2\mu_{k}^2 + \sigtau^2(\Dmat\muvec)_{k}^2\right].
\label{eq:V-matrix}
\end{equation}
The covariance matrices for $\xvec$ and $\yvec$ are then
\begin{equation}
\Sigmax = \Vmat(\muvec), \quad\Sigmay = \Vmat(\psivec).
\label{eq:Sigma-xy}
\end{equation}

The maximum-likelihood (ML) estimate for the model parameters is given by the vectors $\muhatvec$, $\psihatvec$, and $\thetahatvec$ that maximize $L$ subject to the constraint $\psivec = \hmat(\thetavec)\muvec$. Alternatively, we can minimize the negative-log likelihood,
\begin{equation}
\begin{split}
-\ln L(\thetavec, \muvec, \psivec; \xvec, \yvec) &= N\ln(2\pi) + \frac{1}{2}\ln\{\det[\Vmat(\muvec)]\} + \frac{1}{2}\ln\{\det[\Vmat(\psivec)]\} \\
&\quad+ \frac{1}{2}\left\{(\xvec - \muvec)^\transpose\,\left[\Vmat(\muvec)\right]^{-1}(\xvec - \muvec) + (\yvec - \psivec)^\transpose\,\left[\Vmat(\psivec)\right]^{-1}(\yvec - \psivec)\right\},
\end{split}
\label{eq:NLL}
\end{equation}
also subject to the constraint, where the last term now has the familiar least-squares form. The dependence of the covariance matrices on the signal vectors prevents us from using a standard least-squares optimization algorithm to minimize Eq.~(\ref{eq:NLL}), but to a very good approximation we can substitute $\Vmat(\muvec) \approx \Vmat(\xvec)$ and $\Vmat(\psivec) \approx \Vmat(\yvec)$ to obtain
\begin{equation}
\begin{split}
-\ln L(\thetavec, \muvec, \psivec; \xvec, \yvec) &\approx N\ln(2\pi) + \frac{1}{2}\ln\{\det[\Vmat(\xvec)]\} + \frac{1}{2}\ln\{\det[\Vmat(\yvec)]\} \\
&\quad+ \frac{1}{2}\left\{(\xvec - \muvec)^\transpose\,\left[\Vmat(\xvec)\right]^{-1}(\xvec - \muvec) + (\yvec - \psivec)^\transpose\,\left[\Vmat(\yvec)\right]^{-1}(\yvec - \psivec)\right\}.
\end{split}
\label{eq:NLL-approximate}
\end{equation}
The first three terms are now all constant, so we can focus on minimizing the last term, which we multiply by two to obtain a cost function in what is known as a total least-squares form~\cite{vanhuffel1991},
\begin{equation}
\tilde{Q}_\text{TLS}(\thetavec, \muvec, \psivec) = (\xvec - \muvec)^\transpose\,[\Vmat(\xvec)]^{-1}(\xvec - \muvec) + (\yvec - \psivec)^\transpose\,[\Vmat(\yvec)]^{-1}(\yvec - \psivec),
\label{eq:TLS-cost-tilde}
\end{equation}
which is also subject to the constraint $\psivec = \hmat(\thetavec)\muvec$. Note that the total least-squares cost function treats the input and output variables symmetrically, unlike the conventional least-squares cost function in Eq.~(\ref{eq:LS-cost}). Geometrically, it can be interpreted as the sum of the squared distances between each measurement point $(x_k, y_k)$ and its associated model point $(\mu_k, \psi_k)$, using the metric tensors $\Vmat(\xvec)$ and $\Vmat(\yvec)$, respectively, for the input and output variables. As discussed in Sec.~\ref{sec:transfer-function-estimation}, the LS cost function in Eq.~(\ref{eq:LS-cost}) focuses entirely on the deviations in the output variable.

Introducing an $N$-dimensional vector of Lagrange parameters $\lambdavec$ to implement the constraint, we can define the modified cost function,
\begin{equation}
\tilde{\tilde{Q}}_\text{TLS}(\thetavec, \muvec, \psivec, \lambdavec) =  \lambdavec^\transpose\left[\psivec - \hmat(\thetavec)\muvec\right] + \tilde{Q}_\text{TLS}(\muvec, \psivec, \thetavec),
\label{eq:TLS-cost-double-tilde}
\end{equation}
which we may now minimize with respect to $\muvec$, $\psivec$, $\thetavec$, and $\lambdavec$. The parameters of interest are $\thetavec$; minimizing with respect to the remaining parameters gives the equations
\begin{align}
\left.\frac{\partial \tilde{\tilde{Q}}_\text{TLS}}{\partial \lambda_m}\right|_{\muhatvec,\psihatvec,\lambdahatvec,\thetavec} &= \psihat_m - \sum_{l=1}^{N} h_{ml}(\thetavec)\muhat_l = 0,\\
\left.\frac{\partial \tilde{\tilde{Q}}_\text{TLS}}{\partial \psi_m}\right|_{\muhatvec,\psihatvec,\lambdahatvec,\thetavec}
&= \lambdahat_m - 2\sum_{l=1}^{N}\left\{[\Vmat(\yvec)]^{-1}\right\}_{ml}(y_l-\psihat_l) = 0,\\
\left.\frac{\partial \tilde{\tilde{Q}}_\text{TLS}}{\partial \mu_m}\right|_{\muhatvec,\psihatvec,\lambdahatvec,\thetavec} 
&= - \sum_{l=1}^{N}\lambdahat_lh_{lm}(\thetavec) - 2\sum_{l=1}^{N}\left\{[\Vmat(\xvec)]^{-1}\right\}_{ml}(x_l-\muhat_l)  = 0,
\end{align}
which have solutions
\begin{align}
\psihatvec_{\thetavec} &= \hmat(\thetavec)\muhatvec_{\thetavec},\label{eq:psi-hat}\\
\lambdahatvec_{\thetavec} &= 2[\Vmat(\yvec)]^{-1}(\yvec - \psihatvec_{\thetavec}),\label{eq:lambda-hat}\\
\muhatvec_{\thetavec} &= \left\{\bm{1} + \Vmat(\xvec)\left[\hmat(\thetavec)\right]^\transpose[\Vmat(\yvec)]^{-1}\hmat(\thetavec)\right\}^{-1}\left\{\xvec + \Vmat(\xvec)\left[\hmat(\thetavec)\right]^\transpose[\Vmat(\yvec)]^{-1}\yvec\right\}.\label{eq:mu-hat}
\end{align}
Evaluating $\tilde{\tilde{Q}}_\text{TLS}$ at $\muvec = \muhatvec_{\thetavec}$, $\psivec = \psihatvec_{\thetavec}$, and $\lambdavec = \lambdahatvec_{\thetavec}$ yields a simplified cost function that involves only the parameter vector $\thetavec$ and the data vectors $\xvec$ and $\yvec$,
\begin{equation}
Q_\text{TLS}(\thetavec) = (\xvec - \muhatvec_{\thetavec})^\transpose\,[\Vmat(\xvec)]^{-1}(\xvec - \muhatvec_{\thetavec}) + (\yvec - \psihatvec_{\thetavec})^\transpose\,[\Vmat(\yvec)]^{-1}(\yvec - \psihatvec_{\thetavec}).
\label{eq:TLS-cost-xy}
\end{equation}

We can simplify these expressions further by defining
\begin{equation}
 \Umat(\xvec, \thetavec) = \hmat(\thetavec)\Vmat(\xvec)\left[\hmat(\thetavec)\right]^\transpose.
 \label{eq:Uy}
\end{equation}
Substituting Eq.~(\ref{eq:Uy}) in Eqs.~(\ref{eq:mu-hat}) and (\ref{eq:psi-hat}) reveals that $\psihatvec_{\thetavec}$ is just the weighted average of $\yvec$ and $\hmat(\thetavec)\xvec$,
\begin{equation}
\psihatvec_{\thetavec} = \left\{[\Vmat(\yvec)]^{-1} + [\Umat(\xvec,\thetavec)]^{-1}\right\}^{-1}\left\{[\Vmat(\yvec)]^{-1}\yvec + [\Umat(\xvec,\thetavec)]^{-1}[\hmat(\thetavec)\xvec]\right\},
\label{eq:psi-hat-weighted-average}
\end{equation}
where the matrices $\Umat(\xvec,\thetavec)$ and $\Vmat(\yvec)$ are approximately equal to the true (but unknown) covariance matrices $\Cov[\hmat(\thetavec)\xvec] = \Umat(\muvec,\thetavec)$ and $\Cov(\yvec) = \Vmat(\psivec)$, respectively. We emphasize here that although $\hmat(\thetavec)\muvec = \psivec$, the covariance matrices $\Cov[\hmat(\thetavec)\xvec]$ and $\Cov(\yvec)$ are generally different, since $\hmat(\thetavec)$ transforms the noise in $\xvec$ as well as the signal. For example, if $\hmat(A) = A\Imat, A\neq 1,$ and the noise is purely additive, we have $\Cov(\yvec) = \sigalpha^2\Imat\neq\Cov[\hmat(\thetavec)\xvec] = A^2\sigalpha^2\Imat$.

Substituting Eq.~(\ref{eq:Uy}) in Eq.~(\ref{eq:TLS-cost-xy}) yields
\begin{equation}
Q_\text{TLS}(\thetavec) = \left[\yvec - \hmat(\thetavec)\xvec\right]^\transpose\left[\Vmat(\yvec) + \Umat(\xvec, \thetavec)\right]^{-1}\left[\yvec - \hmat(\thetavec)\xvec\right].
\label{eq:TLS-cost-deviations}
\end{equation}
Like the simpler LS cost function in Eq.~(\ref{eq:LS-cost}), the TLS cost function in Eq.~(\ref{eq:TLS-cost-deviations}) is expressed in terms of the deviations $\yvec - \hmat(\thetavec)\xvec$, but is now properly normalized with respect to the associated covariance matrix, $\Vmat(\yvec) + \Umat(\xvec, \thetavec)$. Introducing the matrix square root, $\bm{A}^{1/2} = \bm{X}\leftrightarrow \bm{A} = \bm{X}^2$, we may define the normalized residual vector as
\begin{equation}
\rvec_\text{TLS}(\thetavec) = \left[\Vmat(\yvec) + \Umat(\xvec, \thetavec)\right]^{-1/2}\left[\yvec - \hmat(\thetavec)\xvec\right],
\label{eq:TLS-residuals}
\end{equation}
which allows us to express the TLS cost function in the compact form,
\begin{equation}
Q_\text{TLS}(\thetavec) = [\rvec_\text{TLS}(\thetavec)]^\transpose\rvec_\text{TLS}(\thetavec).
\label{eq:TLS-cost-residuals}
\end{equation}
The TLS estimate for the parameter vector, $\thetahatvec$, may now be determined by minimizing $Q_\text{TLS}(\thetavec)$ using a standard least-squares optimization algorithm.

\begin{figure}[tbp]
\begin{center}
\includegraphics[width=0.9\columnwidth]{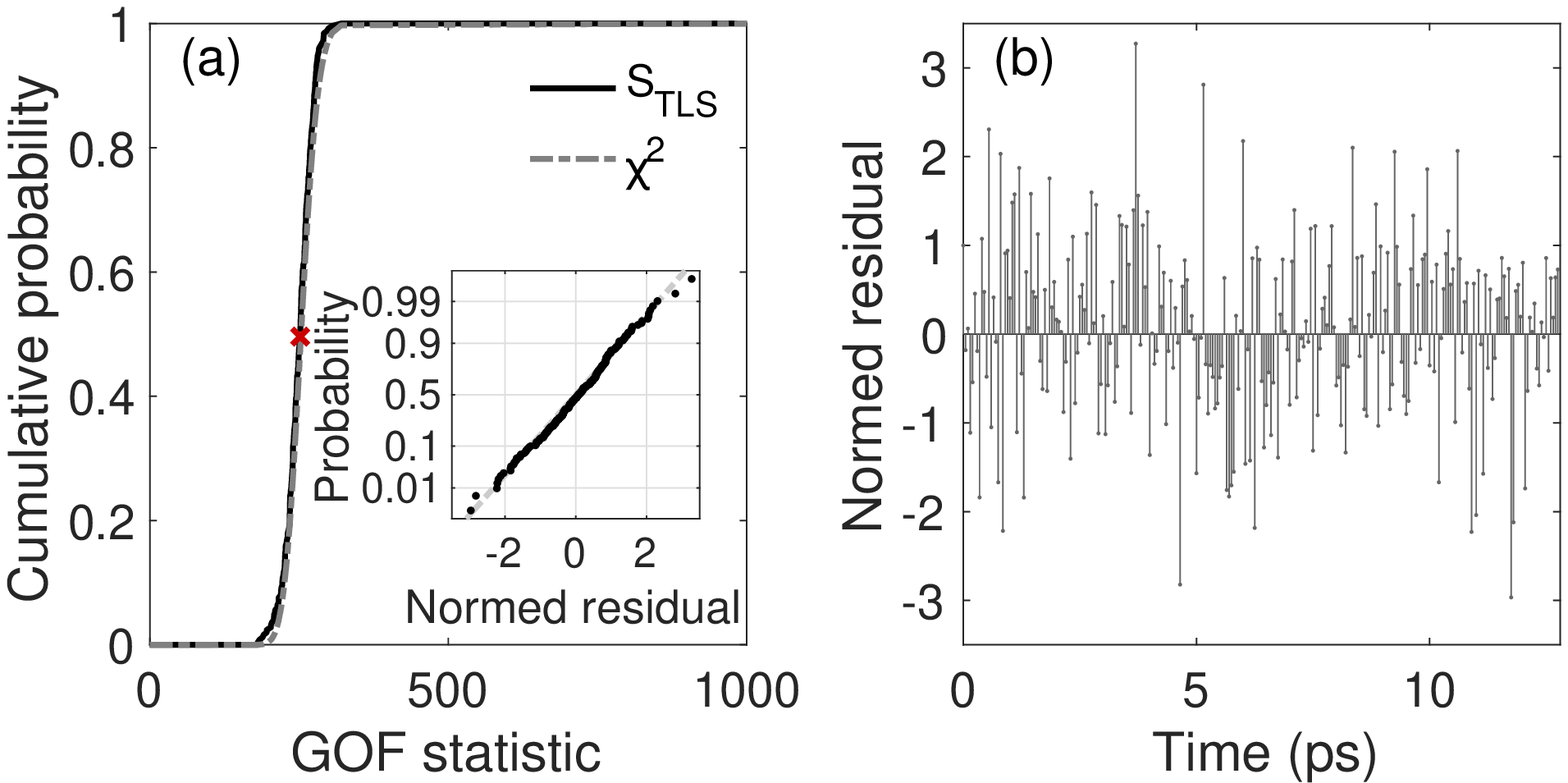}
\caption{Measures of fit quality for TLS fits with Eq.~(\ref{eq:model1}), obtained by minimizing Eq.~(\ref{eq:TLS-cost-residuals}); compare with Fig.~\ref{fig:ETFE-GOF}. (a)~Cumulative distribution of the GOF statistic, $S_\text{TLS}$, for the Monte Carlo simulations shown in Fig.~\ref{fig:ETFE-simulation}. The $\chi^2$ distribution for the same number of degrees of freedom ($\nu = 254$) is shown for comparison. The red $\times$ marks a fit with $S_\text{TLS} \approx 252$, which is just above the median value. The normalized residuals for this fit, $\rvec_\text{TLS}(\thetahatvec)$, are shown in (b) as a function of time, and in the inset to (a) with a normal probability plot (black dots). The gray dash-dotted line in the inset represents the standard normal distribution. }
\label{fig:TLS-GOF}
\end{center}
\end{figure}

Figure~\ref{fig:TLS-GOF} shows fit results for the same model and data shown in Fig.~\ref{fig:ETFE-GOF}, but obtained by minimizing $Q_\text{TLS}(\thetavec)$ in Eq.~(\ref{eq:TLS-cost-residuals}) instead of $Q_\text{ETFE}(\thetavec)$ in Eq.~(\ref{eq:ETFE-cost}). The statistical properties obtained with $Q_\text{TLS}(\thetavec)$ are clearly superior to those with $Q_\text{ETFE}(\thetavec)$. The GOF statistic, $S_\text{TLS} = Q_\text{TLS}(\thetahatvec)$, approximates a $\chi^2$ distribution with $\nu = N-p$ degrees of freedom, as expected. The normalized residual vector $\rvec_\text{TLS}(\thetahatvec)$, shown for one of the fits, exhibits a standard normal distribution with no discernible correlations among neighboring time points. The distribution for $S_\text{TLS}$ is more than 7 times narrower than the distribution for $S_\text{ETFE}$, making it that much more sensitive to a poor fit. Similarly, the lack of structure in $\rvec_\text{TLS}(\thetahatvec)$ makes it more useful for residual analysis than $r_\text{ETFE}(\omega_l; \thetahatvec)$, since structure associated with poor fits may be discerned more readily. Finally, unlike the ETFE fits, the TLS fits yield estimates for the parameter uncertainties that agree with the values observed over the set of Monte Carlo samples, $\sigma_{\theta_1} = 0.002$ and $\sigma_{\theta_2} = 0.4~\text{fs}$, which are both about a factor of 2 smaller than the values observed in the ETFE parameter estimates. In summary, when compared with the standard ETFE fit procedure, the TLS fit procedure offers better discrimination between good models and bad ones, more precise values for the parameters, and more accurate estimates of the parameter uncertainties.

\section{Maximum-likelihood estimation of the noise model}
\label{sec:noise-est}
We have assumed so far that the noise amplitudes $\sigma_\alpha, \sigma_\beta,$ and $\sigma_\tau$ in Eq.~(\ref{eq:V-matrix}) are known, but in general they must also be determined experimentally. One way to do this is to measure the noise at three different points on the THz waveform: as we saw in Fig.~\ref{fig:signal-noise}, the $\sigma_\beta$ term dominates near the peak, the $\sigma_\tau$ term dominates where the signal crosses zero with a large slope, and the $\sigma_\alpha$ term dominates where both the signal and its slope are small. Another approach is to determine the variance as a function of time from a set of repeated waveform measurements~\cite{naftaly2009}, then obtain the noise parameters from a fit with Eq.~(\ref{eq:noise-amplitude}). But both of these methods are susceptible to systematic errors from drift, which produces excess signal variability over the timescale of multiple measurements~\cite{withayachumnankul2008, hubers2011, skorobogatiy2018}. In this section, we describe a likelihood method for estimating the noise parameters that accounts for this drift.

We consider repeated measurements of an ideal primary waveform, $\mu(t)$, which has an amplitude and a delay that drift on a timescale longer than the waveform acquisition time. We can then associate each measurement with an ideal secondary waveform,
\begin{equation}
\zeta(t; A_l, \eta_l) = A_l\mu(t - \eta_l),
\label{eq:zeta}
\end{equation}
where $A_l$ is the relative amplitude, $\eta_l$ is the delay, and $l \in \{0, 1, \ldots, M-1\}$. We also set $A_0 = 1$ and $\eta_0 = 0$, so that $\zeta(t; A_0, \eta_0) = \mu(t)$. As before, we sample these waveforms at the nominal times $t_k = kT, k \in \{0, 1,\ldots, N-1\}$ to obtain the ideal primary waveform vector $\muvec = [\mu_0, \mu_1, \ldots, \mu_{N-1}]^\transpose$ and $M$ measurement vectors $\xvec_l = [x_l(t_0), x_l(t_1), \ldots, x_l(t_{N-1})]^\transpose$, which we can arrange in an $N\times M$ matrix $\Xmat = [\xvec_0, \xvec_1, \ldots, \xvec_{N-1}]$. We also write the amplitudes and delays in vector form, $\Avec = [A_0, A_1, \ldots, A_{M-1}]^\transpose$ and  $\etavec = [\eta_0, \eta_1, \ldots, \eta_{M-1}]^\transpose$, respectively.

With these definitions, we can express the sampled secondary waveforms in vector form,
\begin{equation}
\zetavec_l = \zetavec(\muvec, A_l, \eta_l) = A_l\Smat(\eta_l)\muvec,
\label{eq:zetavec}
\end{equation}
where the matrix $\Smat(\eta_l)$ is defined to impart a delay by $\eta_l$. Using Eq.~(\ref{eq:transfer-matrix}) with the transfer function $H(\omega; \eta) = \exp(i\omega\eta)$, we can write this matrix explicitly as
\begin{equation}
S_{jk}(\eta) = \frac{1}{N}\sum_{l = -N/2+1}^{N/2-1} \exp(i\omega_l\eta) \exp[-2\pi i(j - k)l/N] + \frac{1}{N}\cos(\omega_{N/2}\eta)\cos[\pi(j-k)]
\label{eq:shift-matrix}
\end{equation}
for even $N$, with changes for odd $N$ as described for Eq.~(\ref{eq:transfer-matrix}). Following Eqs.~(\ref{eq:V-matrix}) and (\ref{eq:Sigma-xy}), and arranging the secondary waveform vectors in an $N\times M$ matrix $\Zmat = [\zetavec_0, \zetavec_1,\ldots,\zetavec_{M-1}]$, we also have
\begin{equation}
[\Cov(\xvec_l)]_{jk} = [\Vmat(\zetavec_l; \sigalpha, \sigbeta, \sigtau)]_{jk} = \delta_{jk}\left[\sigalpha^2 + \sigbeta^2 Z_{kl}^2 + \sigtau^2(\Dmat\Zmat)_{kl}^2\right],
\label{eq:covariance-zeta}
\end{equation}
where $\Dmat$ is defined in Eq.~(\ref{eq:D-matrix}), $\zetavec_l$ and $\Zmat$ depend implicitly on $\Avec$ and $\etavec$, and the dependence of $\Vmat$ on the noise amplitudes is now expressed explicitly.

The likelihood function for observing $\Xmat$ under these assumptions is
\begin{equation}
L(\sigma_\alpha,\sigma_\beta,\sigma_\tau,\muvec,\Avec,\etavec;\Xmat) = \prod_{l=0}^{M-1} \frac{\exp\left\{-\frac{1}{2}(\xvec_l - \zetavec_l)^\transpose[\Vmat(\zetavec_l; \sigalpha, \sigbeta, \sigtau)]^{-1}(\xvec_l - \zetavec_l)\right\}}{\sqrt{(2\pi)^N\det[\Vmat(\zetavec_l; \sigalpha, \sigbeta, \sigtau)]}},
\label{eq:likelihood-noise}
\end{equation}
in which the noise amplitudes $\sigalpha$, $\sigbeta$ and $\sigtau$ are the parameters of interest and the signal vectors $\muvec$, $\Avec$ and $\etavec$ are nuisance parameters. As with Eq.~(\ref{eq:likelihood-tf}), it is more convenient computationally to work with the negative-log likelihood,
\begin{equation}
\begin{split}
-\ln[L(\sigma_\alpha,\sigma_\beta,\sigma_\tau,\muvec,\Avec,\etavec;\Xmat)] &= \frac{MN}{2}\ln(2\pi) + \frac{1}{2}\sum_{l=0}^{M-1} \ln\left\{\det[\Vmat(\zetavec_l; \sigalpha, \sigbeta, \sigtau)]\right\} \\
&+ \frac{1}{2}\sum_{l=0}^{M-1}  \left\{(\xvec_l - \zetavec_l)^\transpose[\Vmat(\zetavec_l; \sigalpha, \sigbeta, \sigtau)]^{-1}(\xvec_l - \zetavec_l)\right\}.
\end{split}
\label{eq:NLL-noise}
\end{equation}
Ignoring the constant first term, multiplying the remaining terms by 2, and expressing the matrix elements explicitly, we can define the ML cost function,
\begin{equation}
\begin{split}
Q_\text{ML}(\sigma_\alpha,\sigma_\beta,\sigma_\tau,\muvec,\Avec,\etavec;\Xmat) = \sum_{k=0}^{N-1}\sum_{l=0}^{M-1}& \left\{\ln\left[\sigalpha^2 + \sigbeta^2 Z_{kl}^2 + \sigtau^2(\Dmat\Zmat)_{kl}^2\right]\right. \\
&\quad + \left.\frac{(X_{kl} - Z_{kl})^2}{\sigalpha^2 + \sigbeta^2 Z_{kl}^2 + \sigtau^2(\Dmat\Zmat)_{kl}^2}\right\},
\end{split}
\label{eq:ML-cost}
\end{equation}
which we minimize to obtain ML estimates for all of the free parameters in the model.

The resulting estimates for the noise parameters ($\sigalpha$, $\sigbeta$, and $\sigtau$) are biased below their true values, which we can attribute to the presence of the nuisance parameters ($\muvec$, $\Avec$, and $\etavec$)~\cite{pawitan2001}. For example, in 1000 Monte Carlo simulations of $M=10$ repeated measurements using the waveforms described in Sec.~\ref{sec:signal-noise}, the ratios of the ML estimates to their true values are $\sighatalpha/\sigalpha = 0.95 \pm 0.02$, $\sighatbeta/\sigbeta = 0.94 \pm 0.03$, and $\sighattau/\sigtau = 0.89 \pm 0.09$, all significantly below unity. Increasing the number of observations to $M=50$ reduces the bias to $\sighatalpha/\sigalpha = 0.990 \pm 0.007$, $\sighatbeta/\sigbeta = 0.98 \pm 0.01$, and $\sighattau/\sigtau = 0.93 \pm 0.04$, but some bias remains in $\sighatbeta$ and $\sighattau$ even in the limit $M\rightarrow\infty$. This residual bias arises because the number of elements in $\Avec$ and $\etavec$ both grow with the number of observations, which allows us to account for drift but dilutes some of the information that the data provide about the noise~\cite{pawitan2001,neyman1948}. In principle, we can resolve this problem by integrating out all of the nuisance parameters in Eq.~(\ref{eq:likelihood-noise}) to obtain a marginal likelihood that depends only on $\sigalpha$, $\sigbeta$, and $\sigtau$~\cite{pawitan2001}, but this involves computationally expensive integrations for a relatively small benefit. As we discuss below, we can remove much of the bias by simply rescaling the noise parameters by a suitable correction factor.

To determine the bias correction factor it is helpful to consider a simplified example in which there is no drift and only additive noise, so that $A_l = 1$ and $\eta_l = 0$ for all $l$ and $\sigbeta = \sigtau = 0$. The ML cost function then reduces to
\begin{equation}
\tilde{Q}_\text{ML}(\sigma_\alpha,\muvec;\Xmat) = \sum_{k=0}^{N-1}\sum_{l=0}^{M-1} \left[\ln\sigalpha^2 + \frac{(X_{kl} - \mu_{k})^2}{\sigalpha^2}\right],
\label{eq:ML-cost-simple}
\end{equation}
which is minimized by
\begin{align}
\muhat_k &= \frac{1}{N}\sum_{l = 0}^{M-1} X_{kl} = \bar{x}_k, & \sighatalpha^2 &= \frac{1}{N}\sum_{k = 0}^{N-1}s_M^2(t_k) = \overline{s_M^2},
\label{eq:ML-est-simple}
\end{align}
where
\begin{equation}
s_M^2(t_k) = \frac{1}{M}\sum_{l=0}^{M-1}(X_{kl} - \bar{x}_k)^2.
\label{eq:samplevar}
\end{equation}
This result for $\sighatalpha^2$ is biased because it is the waveform average of $s_M^2(t_k)$, which in turn is just the biased sample variance of the measurements at $t_k$. To remove this bias we can apply Bessel's correction, which replaces the factor of $1/M$ with $1/(M-1)$ in Eq.~(\ref{eq:samplevar}). Alternatively, we can multiply $\sighatalpha^2$ by $M/(M-1)$ in Eq.~(\ref{eq:ML-est-simple}). Returning to Eq.~(\ref{eq:ML-cost}) and following similar reasoning, we can justify rescaling all of the ML noise parameter estimates by the same factor,
\begin{align}
\sighatalpha^* &= \sighatalpha\sqrt{\frac{M}{M-1}}, & \sighatbeta^* &= \sighatbeta\sqrt{\frac{M}{M-1}}, & \sighattau^* &= \sighattau\sqrt{\frac{M}{M-1}}.
\label{eq:unbiasedsigma}
\end{align}
With these corrections, the Monte Carlo simulations with $M=10$ yield $\sighatalpha^*/\sigalpha = 1.00 \pm 0.02$, $\sighatbeta^*/\sigbeta = 0.99 \pm 0.03$, and $\sighattau^*/\sigtau = 0.94 \pm 0.09$, which are all within the statistical uncertainty of the true values. The simulations with $M=50$ yield lower statistical uncertainty, but with the same average values: $\sighatalpha^*/\sigalpha = 1.000 \pm 0.007$, $\sighatbeta^*/\sigbeta = 0.99 \pm 0.01$, and $\sighattau^*/\sigtau = 0.94 \pm 0.04$. For all practical purposes, the remaining bias in $\sighatbeta^*$ and $\sighattau^*$ is negligible.

\section{Experimental verification}
\label{sec:Verification}

\begin{figure}[tbp]
\begin{center}
\includegraphics[width=0.9\columnwidth]{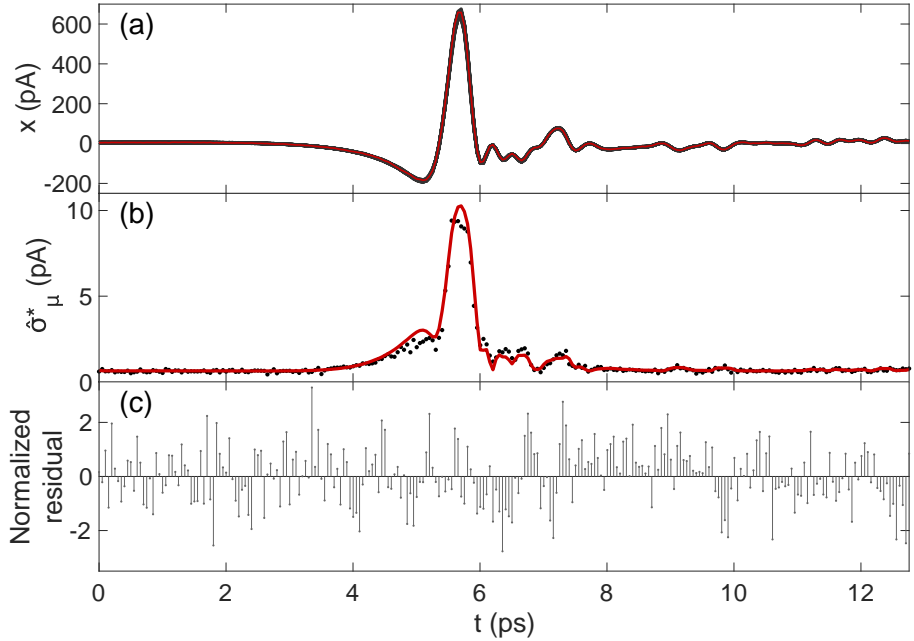}
\caption{Noise-model fits with laboratory measurements. (a) Set of 50 nominally identical waveform measurements that compose the data matrix $\Xmat$. (b) Time-dependent noise amplitude obtained from a fit with Eq.~(\ref{eq:ML-cost}) to $\Xmat$. The solid line shows the ideal noise model, with the bias-corrected ML estimates $\sighatalpha^* = (0.623\pm 0.005)~\text{pA}$, $\sighatbeta^* = (1.55\pm 0.03)\text{\%}$, and $\sighattau^* = (1.8\pm 0.1)~\text{fs}$. Points show the standard deviation of the measurements after correcting for drift, as described in the text. (c) Normalized residuals for one of the waveform measurements shown in (a).}
\label{fig:noise-model}
\end{center}
\end{figure}

In this section we present experimental results that verify our analysis. Figure~\ref{fig:noise-model}(a) shows the raw data that we use to specify the noise model, $\Xmat$, which comprises $M=50$ waveforms, each with $N=256$ points sampled every $T = 50~\text{fs}$. With this data as input, we minimize Eq.~(\ref{eq:ML-cost}) to obtain the ML estimates, $\sighatalpha$, $\sighatbeta$, $\sighattau$,  $\hat{\muvec}$, $\hat{\Avec}$, and $\hat{\etavec}$ for all of the free parameters in the model. The resulting time-dependent noise amplitude, corrected for bias using Eq.~(\ref{eq:unbiasedsigma}), is
\begin{equation}
\hat{\sigma}^*_\mu(t_k) = \sqrt{V_{kk}(\muhatvec; \sighatalpha^*, \sighatbeta^*, \sighattau^*)}.
\label{eq:ML-noise-amplitude}
\end{equation}
In Fig.~\ref{fig:noise-model}(b) we compare this model to the observed time-dependent noise, which we estimate by using $\hat{\Avec}$ and $\hat{\etavec}$ to correct for the drift,
\begin{equation}
\tilde{\xvec}_l = \frac{1}{\hat{A}_l}\Smat(-\hat{\eta}_l)\xvec_l,
\label{eq:X-corrected}
\end{equation}
then compute the unbiased standard deviation at each time point over the set $\{\tilde{\xvec}_l\}$. The model quantitatively describes most features of the measurements, with small deviations near some of the peaks. As a further consistency check, Fig.~\ref{fig:noise-model}(c) shows the normalized residuals for one of the waveforms,
\begin{equation}
\rvec_{\text{ML},l} = \left\{\Vmat\left[\zetavec(\hat{\muvec}, \hat{A}_l, \hat{\eta}_l); \sighatalpha^*, \sighatbeta^*, \sighattau^*\right]\right\}^{-1/2}\left[\xvec_{l} - \zetavec(\hat{\muvec}, \hat{A}_l, \hat{\eta}_l)\right],
\end{equation}
which demonstrates that they approximately follow a standard normal distribution, with small but statistically significant correlations among neighboring points.

\begin{figure}[tbp]
\begin{center}
\includegraphics[width=0.9\columnwidth]{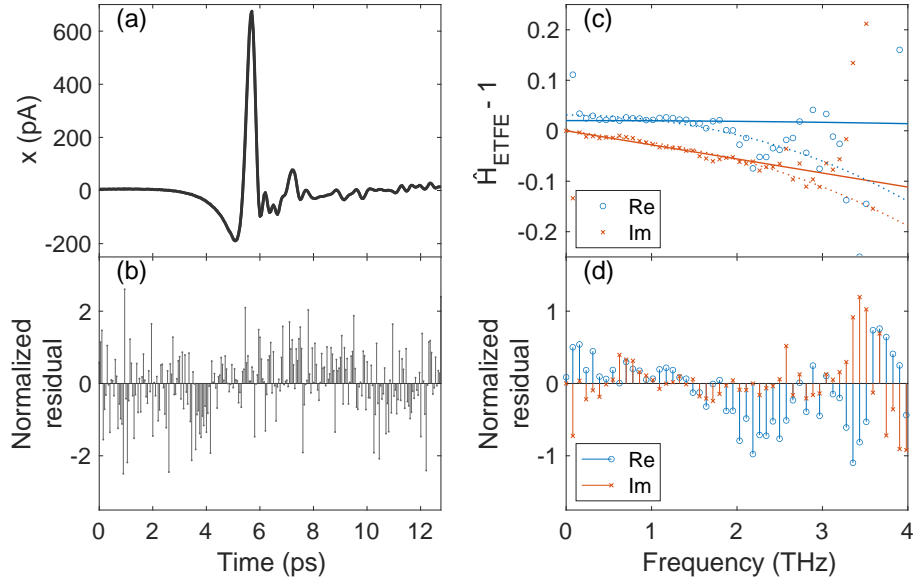}
\caption{Transfer-function fits with laboratory measurements. (a) Two sequential waveform measurements, taken from the set shown in Fig.~\ref{fig:noise-model}. (b) Normalized time-domain residuals, $\bm{r}_\text{TLS}(\thetahatvec^\text{TLS})$, for the TLS fit with Eq.~(\ref{eq:model1}). (c) Real and imaginary parts of $\hat{H}_\text{ETFE}(\omega_l) - 1$ (dots), fitted with Eq.~(\ref{eq:model1}) (solid lines) and Eq.~(\ref{eq:model2}) (dotted lines) by minimizing $Q_\text{ETFE}$ in Eq.~(\ref{eq:ETFE-cost}). (d) Normalized frequency-domain residuals, $r_\text{ETFE}(\omega_l; \thetahatvec^\text{ETFE})$, for the fit with Eq.~(\ref{eq:model1}).}
\label{fig:TF-fits}
\end{center}
\end{figure}

Figure~\ref{fig:TF-fits} shows fits to two sequential waveforms from this set. A fit with Eq.~(\ref{eq:model1}) in the time domain, obtained by minimizing $Q_\text{TLS}$ in Eq.~(\ref{eq:TLS-cost-deviations}), yields $\thetahat^\text{TLS}_1 = 1.019\pm 0.003$ for the amplitude and $\thetahat^\text{TLS}_2 = (-2.8\pm 0.5)~\text{fs}$ for the delay, with the normalized residuals shown in Fig.~\ref{fig:TF-fits}(b). A fit with the same model in the frequency domain, obtained by minimizing $Q_\text{ETFE}$ in Eq.~(\ref{eq:ETFE-cost}), yields $\thetahat^\text{ETFE}_1 = 1.020\pm 0.004$ and $\thetahat^\text{ETFE}_2 = (-4.4\pm 0.6)~\text{fs}$, with the normalized residuals shown in Fig.~\ref{fig:TF-fits}(d). Both fit methods reveal significant differences from the ideal values of $\theta_1 = 1$ and $\theta_2 = 0$, reflecting the measurement drift discussed in Sec.~\ref{sec:noise-est}. As we found for the residuals of the noise-model fit in Fig.~\ref{fig:noise-model}(c), the residuals of the transfer-function fit in Fig.~\ref{fig:TF-fits}(b) show small but statistically significant correlations. But as we also found with the idealized Monte Carlo simulations, the residuals of the ETFE fit in the frequency domain are much more structured than the residuals of the TLS fit to the same data in the time domain. 

To illustrate the analysis problem that this raises, in Fig.~\ref{fig:TF-fits}(c) we compare an ETFE fit with Eq.~(\ref{eq:model1}) to an ETFE fit with a more flexible transfer-function model,
\begin{equation}
H_2(\omega; \thetavec) = (\theta_1 + \theta_3\omega^2)\exp[i\omega(\theta_2 + \theta_4\omega^2)].
\label{eq:model2}
\end{equation}
Since the two input waveforms are nominally identical, we know that the downturns in $\Re[\hat{H}_\text{ETFE}(\omega)]$ and $\Im[\hat{H}_\text{ETFE}(\omega)]$ with increaing frequency near 2~THz are spurious. But if we did not know this in advance, we might conclude from Fig.~\ref{fig:TF-fits}(c) that Eq.~(\ref{eq:model2}) describes the measurements better than Eq.~(\ref{eq:model1}). We would also be able to support this conclusion by comparing the GOF statistic for the fit with Eq.~(\ref{eq:model1}), $S^{(1)}_{\text{ETFE}} \approx 223$, with $\nu = 254$ degrees of freedom, to that obtained with Eq.~(\ref{eq:model2}), $S^{(2)}_{\text{ETFE}} \approx 191$, with $\nu = 252$ degrees of freedom. By adding only two additional fit parameters, we reduce $S_\text{ETFE}$ by 33, which erroneously suggests that the added complexity of Eq.~(\ref{eq:model2}) captures a real physical effect. The TLS method is more robust against such overfitting: the GOF statistic with Eq.~(\ref{eq:model1}) is $S^{(1)}_{\text{TLS}} \approx 198$, while with Eq.~(\ref{eq:model2}), $S^{(2)}_{\text{TLS}} \approx 196$. In this case, adding two free parameters reduces $S_\text{TLS}$ by only two, so Occam's razor favors the simpler model, Eq.~(\ref{eq:model1}). More formally, to select from a set of transfer-function models $H_k(\omega; \thetavec_k)$, $k = 1, 2, \ldots, N_\text{model}$, each with $p_k$ free parameters, we can minimize a modified cost function based on the Akaike information criterion~\cite{pawitan2001},
\begin{equation}
Q_{\AIC}(k) = S_\text{TLS}^{(k)} + 2p_k,
\label{eq:AIC}
\end{equation}
where $S_\text{TLS}^{(k)}$ is the TLS GOF statistic for the model $H_k (\omega; \thetavec_k)$. As we discussed in Sec.~\ref{sec:MLE-TF}, this ability to discriminate among competing models is one of the major advantages of the TLS method.

If we divide the 50 experimental waveforms into 25 sequential pairs and fit each pair with Eq.~(\ref{eq:model1}), the results are qualitatively consistent with the Monte Carlo simulations and support the conclusion that $S_\text{TLS}$ offers the better absolute measure of fit quality. The distribution of $S_\text{ETFE}$ over the experiments has a mean $\bar{S}_\text{ETFE} \approx 235$ and standard deviation $\sigma(S_\text{ETFE}) \approx 113$, while the distribution for $S_\text{TLS}$ has a mean $\bar{S}_\text{TLS} \approx 246$ and standard deviation $\sigma(S_\text{TLS}) \approx 39$. For the simulated distributions shown in Fig.~\ref{fig:ETFE-GOF}(a) and Fig.~\ref{fig:TLS-GOF}(a), we obtain $\bar{S}_\text{ETFE} \approx 235$, $\sigma(S_\text{ETFE}) \approx 160$, $\bar{S}_\text{TLS} \approx 250$, and $\sigma(S_\text{TLS}) \approx 22$. As we discussed at the end of Sec.~\ref{sec:MLE-TF}, a narrower GOF distribution provides better sensitivity to fit quality. And while the experimental distribution for $S_\text{TLS}$ is nearly twice as broad it is in the simulations, it is still nearly a factor of 3 narrower than the experimental distribution for $S_\text{ETFE}$.  Despite the quantitative differences between the simulations and the experiment, the TLS method consistently shows better performance than the ETFE.

\section{Conclusion}
\label{sec:Conclusion}
In summary, we have developed a maximum-likelihood approach to THz-TDS analysis and demonstrated that it has numerous advantages over existing methods. Starting from a few simple assumptions, we derived a method to determine the transfer function relationship between two THz-TDS measurements, together with a method to specify a parameterized noise model from a set of repeated measurements. In each case, we also derived expressions for fit residuals in the time domain that are properly normalized by the expected noise. Through a combination of Monte Carlo simulations and experimental measurements, we verified that these tools yield results that are accurate, precise, responsive to fit quality, and generally superior to the results of fits to the ETFE.

We focused on simple examples to emphasize the logical structure of the method, but we can readily apply the same approach to more complex problems. For example, we have successfully used the framework presented here to measure a weak frequency dependence in the Drude scattering rate of a metal, which is predicted by Fermi liquid theory; we have also used it to measure small variations in the THz-frequency refractive index of silicon with temperature~\cite{mohtashemi2020}. In both of these applications we found that maximum-likelihood analysis in the time domain provided better performance than a least-squares analysis based on the ETFE in the frequency domain. We expect similar improvements in other applications, and provide MATLAB functions in Code Repository 1 (Ref.~\cite{dodge2020}) for others to try.

\begin{backmatter}
\bmsection{Funding}
JSD acknowledges support from NSERC and CIFAR, and DGS from an NSERC Alexander Graham Bell Canada Graduate Scholarship.

\bmsection{Acknowledgments}
JSD thanks John Bechhoefer for encouragement and stimulating discussions as this work developed.

\bmsection{Disclosures}
The authors declare no conflicts of interest.
\end{backmatter}

\bibliography{thztdmle}

\end{document}